\newcommand*{\ditto}{\raisebox{-0.5ex}{\ttfamily"}}
\documentclass[letterpaper]{aastex}
\usepackage{amsbsy}
\usepackage{amsfonts}
\usepackage{amssymb}
\usepackage{lscape}
\usepackage{graphicx}% Include  files
\usepackage{graphics}
\usepackage{bm}% bold math
\begin{document}
%\preprint{APS/123-QED}

\title{Influence of Electron-Impact Multiple Ionization on Equilibrium and Dynamic Charge State Distributions: A Case Study Using Iron}

\author{M. Hahn\altaffilmark{1} and D. W. Savin\altaffilmark{1}}

\altaffiltext{1}{Columbia Astrophysics Laboratory, Columbia University, 550 West 120th Street, New York, NY 10027 USA}
%\altaffiltext{2}{Department of Physics, University of Strathclyde, 107 Rottenrow East, Glasgow G4 0NG, UK}
%\altaffiltext{3}{Max-Planck-Institut f\"{u}r Kernphysik, Saupfercheckweg 1, 69117 Heidelberg, Germany}
%\altaffiltext{4}{GSI Helmholtzzentrum f\"ur Schwerionenforschung, Planckstrasse 1, 64291 Darmstadt, Germany }
%\altaffiltext{5}{Institut f\"{u}r Atom- und Molek\"{u}lphysik, Justus-Liebig-Universit\"{a}t Giessen, Leihgesterner Weg 217, 35392 Giessen, Germany}

%\begin{center}
\date{\today}% It is always \today, today,
%\end{center}
\begin{abstract}
	
	We describe the influence of electron-impact multiple ionization (EIMI) on the ionization balance of collisionally ionized plasmas. We are unaware of any previous ionization balance calculations that have included EIMI, which is usually assumed to be unimportant. Here, we incorporate EIMI cross-section data into calculations of both equilibrium and non-equilibrium charge-state distributions (CSDs). For equilibrium CSDs, we find that EIMI has only a small effect and can usually be ignored. However, for non-equilibrium plasmas the influence of EIMI can be important. In particular, we find that for plasmas in which the temperature oscillates there are significant differences in the CSD when including versus neglecting EIMI. These results have implications for modeling and spectroscopy of impulsively heated plasmas, such as nanoflare heating of the solar corona.
		
\end{abstract}
	
\maketitle
	
\section{Introduction}\label{sec:intro}

	Collisionally ionized plasmas are formed in numerous astrophysical sources, such as the Sun and other stars, supernova remnants, galaxies, and the intracluster medium of galaxy clusters. Interpreting observations and modeling astrophysical processes in these objects requires knowledge of the underlying charge state distribution (CSD) of the plasma. The CSD is determined by the rates of ionization and recombination. In collisionally ionized plasmas, the ions are ionized by electron-impact ionization (EII). The electron-ion recombination is dominated by dielectronic recombination (DR) and radiative recombination (RR). These various processes have been reviewed by \citet{Muller:Book:2008}. 

	An electron-ion collision can also cause electron-impact multiple ionization (EIMI) --- the ejection of multiple electrons due to a single collision. This process has generally been ignored in ionization balance calculations because, for a given charge state, multiple ionization usually becomes significant only at temperatures so high that the fractional abundance of that charge state is small \citep{Tendler:PhysLett:1984}. However, it has been argued that multiple ionization may become important in dynamic systems where the ions are suddenly exposed to higher electron temperatures \citep{Muller:PhysLett:1986}. Such non-equilibrium could occur, for example, in solar flares \citep{Reale:ApJ:2008,Bradshaw:ApJS:2011}, supernova remnants \citep{Patnaude:ApJ:2009}, or merging galaxy clusters \citep{Akahori:PASJ:2010}. Nevertheless, calculations of these dynamic events have ignored multiple-ionization processes. 
	
	One reason for not considering multiple ionization, is that very little EIMI data exist. To theoretically calculate even double-ionization cross sections is very difficult. This is because the problem requires considering at least four charged particles in the outgoing channel: the ion, the colliding electron, and at least two ejected electrons. The interactions among all these particles via the Coulomb potential must be accounted for \citep{Berakdar:PhysLett:1996,Gotz:JPhysB:2006}. Thus, most multiple-ionization cross sections are from experimental measurements or semiempirical formulae derived from fits to experimental data. To calculate the CSD necessary for astrophysics, data are needed for essentially all the charge states of all the elements from H--Zn. Given practical limitations, experimental studies cannot generate all these required data. 
	
	For a few systems, however, there now exist sufficient empirical data to incorporate multiple ionization into CSD calculations. Here we focus on iron, which is a cosmically abundant element that forms many emission lines commonly used in astrophysical spectroscopy. We have used semiempirical formulae calibrated to experimental data to derive multiple-ionization cross sections for the various charge states of iron. These cross sections were then incorporated into CSD calculations for both equilibrium and dynamic plasmas. Our results confirm that for equilibrium plasmas, multiple ionization has little effect on the charge balance, modifying the ion abundances by at most about 5\%. Conversely, in evolving plasmas, the effects can be significant. 
	
	The rest of this paper is organized as follows. In Section~\ref{sec:cross} we discuss the experimental data sources and semiempirical fitting formulae used for our EIMI cross sections. Section~\ref{sec:csd} briefly reviews the relation between the cross sections and the CSD. In Section~\ref{sec:equilib} we present our ionization-equilibrium calculations and compare them to calculations that consider only electron-impact single ionization (EISI). Section~\ref{sec:dynamic} explores the influence of EIMI on the CSD in situations where the temperature varies rapidly. Our conclusions are summarized in Section~\ref{sec:conclusions}. 
	
\section{Electron-Impact Multiple-Ionization Cross Sections}\label{sec:cross}

	%Describe Multiple Ionization processes - direct double ionization, excitation auto double ionization, ionization autoionization. 

	EIMI involves the ejection of two or more electrons from an ion following a single collision. There are a number of specific processes leading to multiple ionization, many of which are analogous to those for EISI \citep[][]{Muller:Book:2008}.  A collision can result in direct ionization of two or more electrons starting at the multiple-ionization threshold, which depends on the number of electrons to be removed. Excitation-multiple-autoionization (EMA) occurs when an electron excites an electron in the target to a level that decays by ejecting two or more electrons. For highly charged ions, the dominant multiple-ionization process is often ionization-autoionization (IA), in which a collision directly ionizes a core electron and additional electrons are released when system relaxes to fill the resulting hole \citep[e.g.,][]{Muller:PRL:1980,Cherkani:PhysScr:2001,Hahn:ApJ:2011}. 

\subsection{Fitting Formulae}\label{subsec:form}

	Because quantum-mechanical calculations for multiple-ionization cross sections are challenging, EIMI cross sections are generally estimated using semiempirical formulae. Here, we review several semiempirical formulae that have been proposed to model EIMI cross sections. Later we apply these formulae to EIMI cross sections for iron ions. 
	
	\citet{Shevelko:PhysScr:1995,Shevelko:JPhysB:1995} presented a semiempirical formula that describes cross sections for direct multiple ionization of at least three electrons, i.e., triple ionization, from neutral and ionic targets. \citet{Belenger:JPhysB:1997} extended their formulae to double ionization. The cross sections are approximated by:  
\begin{equation}
\sigma_{\mathrm{D}}=\frac{p_{0} p_{1}^{p_{2}}}{\left(E_{\mathrm{th}}/E_{\mathrm{Ryd}}\right)^2}\left(\frac{u+1}{u}\right)^{p_{3}}\frac{\ln{(u)}}{u} 
\times 10^{-18} \,\mathrm{cm^{2}},
\label{eq:shevtar}
\end{equation}
where $u=E/E_{\mathrm{th}}$ is the incident electron energy $E$ normalized by the multiple-ionization threshold $E_{\mathrm{th}}$ for the process being considered and $E_{\mathrm{Ryd}}=13.606$~eV. Here $p_{0}$ and $p_{2}$ are parameters that depend on the number of electrons being removed and have been tabulated by \citet{Shevelko:JPhysB:1995} and \citet{Belenger:JPhysB:1997}. The parameter $p_{1}$ is the number of electrons in the target ion and $p_{3}=1.0$ for neutral targets or $0.75$ for ionic targets. The fits were performed to experimental measurements of ionization from neutral and relatively low charged ions, for which the dominant EIMI process is direct ionization. Thus, this formula is reasonable for describing direct multiple ionization, but other contributions must be added in for systems in which IA becomes important. 

	\citet{Shevelko:JPhysB:2005} presented a semiempirical double ionization formula that accounts for both direct ionization and IA for initially He-like to Ne-like ions. The total cross section is a sum of a direct and an IA cross section. The direct cross section is given by 
\begin{equation}
\sigma_{\mathrm{D}}=1-\mathrm{e}^{-3\left(u-1\right)}\left\{ \frac{p_0}{E_{\mathrm{th}}^3}\left[\frac{u-1}{\left(u+0.5\right)^2}\right] \right\}
\times10^{-13} \, \mathrm{cm^{2}}.
\label{eq:shev5a}
\end{equation}
Here $p_0$ is a tabulated fitting parameter that varies depending on the initial isoelectronic sequence of the ion. $E_{\mathrm{th}}$ is again the threshold for the ionization process being considered, i.e., here direct EIMI. The indirect cross section, due to IA, is given by
\begin{equation}
\sigma_{\mathrm{IA}}=f \frac{p_{0}}{E_{\mathrm{th}}^2}\frac{u-1}{u \left(u+p_{1}\right)} 
\times10^{-13} \, \mathrm{cm^{2}},
\label{eq:shev5b}
\end{equation}
where $E_{\mathrm{th}}$ is the relevant threshold for the process, i.e., the threshold for single ionization of a core electron. The parameters $p_{0}$ and $p_{1}$ depend on the isoelectronic sequence of the initial ion configuration. The quantity $f$ is the branching ratio for autoionization of the intermediate state that is missing an inner-shell electron. \citet{Shevelko:JPhysB:2005} have calculated these branching ratios for K-shell vacancies having configurations $1s 2s^2$ through $1s 2s^2 2p^6 3s$ for nuclear charges $Z=3$--$26$. 

	For certain ions where experimental data exist, \citet{Shevelko:JPhysB:2006} presented a more accurate formula for double-ionization cross sections. . Here, the cross section is the sum of a direct ionization cross section and several possible indirect ionization channels. In this scheme the direct cross section is given by 
\begin{equation}
\sigma_{\mathrm{D}}=1-\mathrm{e}^{-3\left(u-1\right)} \frac{p_0}{E_{\mathrm{th}}^3}\left[\frac{u-1}{\left(u+0.5\right)^2}\right]\left[1+0.1\ln{(4u+1)}\right] 
\times10^{-13} \, \mathrm{cm^{2}},
\label{eq:shev6a}
\end{equation}
where $p_0$ is a fit parameter. The individual indirect ionization cross sections sections, due to IA, are given by 
\begin{equation}
\sigma_{\mathrm{IA}}=\frac{p_0}{E_{\mathrm{th}}^2}\frac{u-1}{u \left(u+5.0 \right)} \left[1+\frac{0.3}{p_{1}}\ln{(4u+1)}\right]
\times10^{-13} \, \mathrm{cm^{2}},
\label{eq:shev6b}
\end{equation}
where $p_0$ is a fit parameter and $p_1$ is the principal quantum number of the core electron that is directly ionized. Unlike Equations~(\ref{eq:shev5a}) and (\ref{eq:shev5b}), for Equations~(\ref{eq:shev6a}) and (\ref{eq:shev6b}) the parameters do not have a tabulated dependence on an isoelectronic sequence, but rather are determined by fits to experimental measurements. Thus, the formulae should be accurate, but it is not possible to extrapolate them to systems that have not been measured. 

	For highly charged ions, the indirect IA process is the dominant contribution to the EIMI cross section. In such cases, the cross section is often well described by multiplying the \citet{Lotz:ZPhys:1969} formula for single ionization of the core electron by the branching ratio $f$ for autoionization of the intermediate state, giving
\begin{equation}
\sigma_{\mathrm{IA}}=4.5 f p_{0}\frac{\ln{(u)}}{E_{\mathrm{th}}^{2}u}
\times10^{-14} \, \mathrm{cm^{2}}.
\label{eq:lotz}
\end{equation}
Here $p_0$ is the initial number of electrons in the level where the ionization takes place. The branching ratios for many ions of astrophysical interest have been given by \citet{Kaastra:AAS:1993} and updated calculations for some systems have been given by \citet{Bautista:AA:2003}, \citet{Gorczyca:ApJ:2003,Gorczyca:ApJ:2006}, \citet{Palmeri:AA:2003}, \citet{Mendoza:AA:2004}, and \citet{Shevelko:JPhysB:2005}. 

\subsection{Application of Fitting Formulae to Iron Ions}\label{subsec:appfe}

	In order to determine the EIMI cross sections for our CSD calculations, we have selected the semiempirical formulae for each system that seem to best reproduce the experimental measurements. Experimental data are not available for every system, so in some cases we have had to estimate the cross section based on the results for nearby charge states. Table~\ref{table:cross} lists the parameters used for each cross section and also indicates for which data there are experimental results. We denote the initial charge state as $q_{i}$ and the final charge state as $q_{f}$. The total cross section is the sum of the individual cross sections for each ionization channel. That is,
\begin{equation}
\sigma_{\mathrm{total}}=\sigma_{\mathrm{D}}+\sum_{N}{\sigma_{\mathrm{IA},i}},
\label{eq:sigtot}
\end{equation}
where $\sigma_{\mathrm{D}}$ is the direct ionization cross section and the $\sigma_{\mathrm{IA},i}$ represent the IA sections arising from $N$ different IA channels, such as due to holes in different shells. Other indirect ionization channels, such as EMA, have been neglected. We note, though that in many cases these other indirect contributions are, fortuitously, roughly accounted for in the semiempirical formulae. This is because the fitting parameters used in the formulae were usually estimated without attempting to distinguish direct and indirect channels, except for IA. Below, we discuss in more detail the experimental measurements and formulae used for each EIMI cross section in our calculations. 

	In the following, the direct EIMI thresholds are generally from \citet{NIST:2013}, while the thresholds for ionization of core electrons come from \citet{Kaastra:AAS:1993} unless otherwise noted. The ionization thresholds for $K$-shell ionization of a core electron are an order of magnitude greater than any other relevant thresholds, being $\gtrsim7000$~eV. Thus $K$-shell IA contributes significantly to the EIMI rate coefficient at temperatures $\gtrsim10^{8}$~K. In most cases, these contributions are included using the Lotz-formula scaled by the branching ratios, i.e., Equation~(\ref{eq:lotz}). However, it is worth noting that this approximation ignores relativistic effects, which may become important at such high energies. In particular, at high energies the Lotz formula cross section falls off like $\ln(u)/u$ as predicted by the Bethe approximation, but in the relativistic limit the Bethe-approximation cross section becomes a constant \citep{Sampson:PhysRep:2009}. %Since we consider temperatures below 10$^{8}$~K, we ignore $K$-shell ionization.

\begin{figure}
\centering \includegraphics[width=0.9\textwidth]{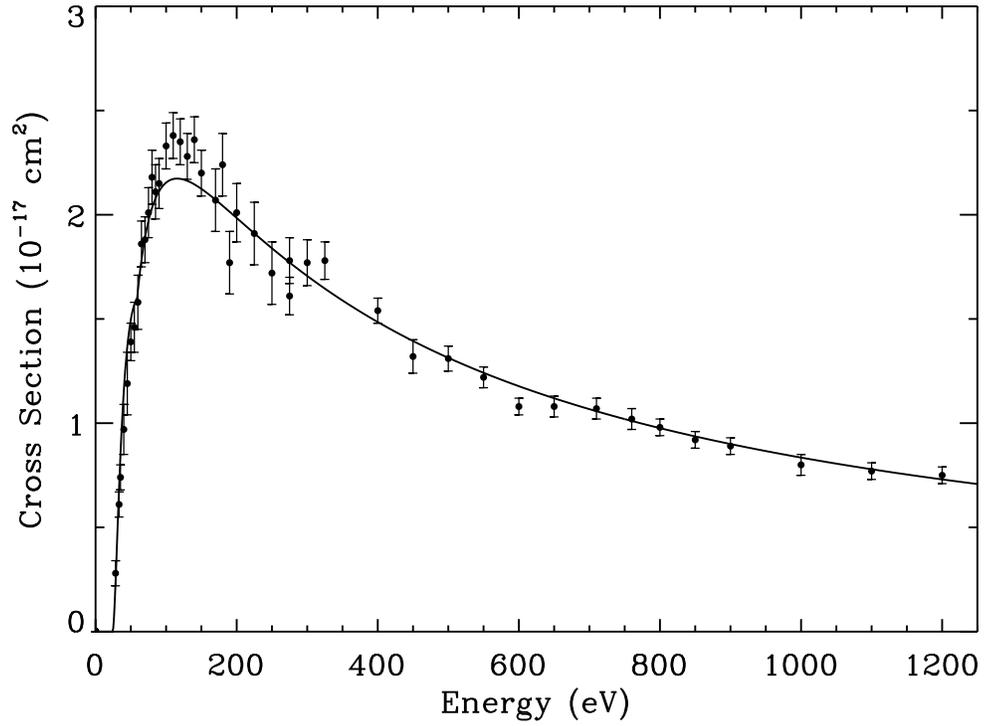}
\caption{\label{fig:fe02} Double ionization of Fe$^{0+}$ forming Fe$^{2+}$. The data points show the experimental results of \citet{Shah:JPhysB:1993} and the solid curve illustrates the fit from \citet{Shevelko:JPhysB:2006}, which we use here.
}
\end{figure}

	Experimental measurements of EIMI starting from Fe$^{0+}$, including double ($q_f=2$), triple ($q_f=3$), and quadruple ($q_f=4$) ionization, were measured by \citet{Shah:JPhysB:1993}. Double ionization of Fe$^{0+}$ was also measured earlier by \citet{Freund:PRA:1990}. However, those measurements seem to be affected by metastable levels in the atom beam. So, we base our cross sections on the data from \citet{Shah:JPhysB:1993}. For double ionization, the parameters given in Table~\ref{table:cross} were derived by \citet{Shevelko:JPhysB:2006} by performing a least squares fit to the experimental data using Equations~(\ref{eq:shev6a}) and (\ref{eq:shev6b}). Figure~\ref{fig:fe02} illustrates this cross section compared to the experimental data. We found that the triple-ionization data were well fit by scaling Equation~(\ref{eq:shevtar}), which describes the direct ionization, and then accounting for the indirect ionization channels by adding to that the Lotz cross sections for the $3p$ and $3s$ ionization multiplied by the appropriate branching ratios from \citet{Kaastra:AAS:1993}. The quadruple-ionization cross section was well described by the fit to Equation~(\ref{eq:shevtar}) given by \citet{Shevelko:JPhysB:1995}. As there are no experimental data for higher order ionization processes ($q_f \geq 5$), we estimate those cross sections by assuming only indirect contributions described by the Lotz formula scaled by the branching ratios from \citet{Kaastra:AAS:1993}. Similarly, there are no experimental data at the energies relevant for $K$-shell IA, so we also estimate those cross sections in the same way. 
	
	Double-ionization cross sections for Fe$^{1+}$ and Fe$^{3+}$--Fe$^{6+}$ were measured by \citet{Stenke:JPhysB:1999}. Fits to these data were given by \citet{Shevelko:JPhysB:2006} using Equations~(\ref{eq:shev6a}) and (\ref{eq:shev6b}). We use their fits here. The existing experimental measurements do not extend high enough to benchmark $K$-shell IA, so here we have used the Lotz formula and the branching ratios given by \citet{Kaastra:AAS:1993} to incorporate those processes. Similarly, for higher order EIMI of these systems we assume that the dominant process is IA and estimate the cross section using the Lotz formula and the branching ratios. 
	
	There are no experimental data for double ionization of Fe$^{2+}$. However, Ni$^{4+}$ is isoelectronic and double ionization of this ion was measured by \citet{Stenke:NIMB:1995} and was the basis for a semiempirical fit reported in \citet{Shevelko:JPhysB:2006}. We find that the Ni$^{4+}$ double-ionization cross section can be reproduced well using Equation~(\ref{eq:shevtar}) for the direct ionization contribution plus scaled Lotz cross sections for the indirect contributions. Thus, we have applied the same formulae to the Fe$^{2+}$ double-ionization cross section, with appropriate modifications for the different energy thresholds and branching ratios. 
	
\begin{figure}
\centering \includegraphics[width=0.9\textwidth]{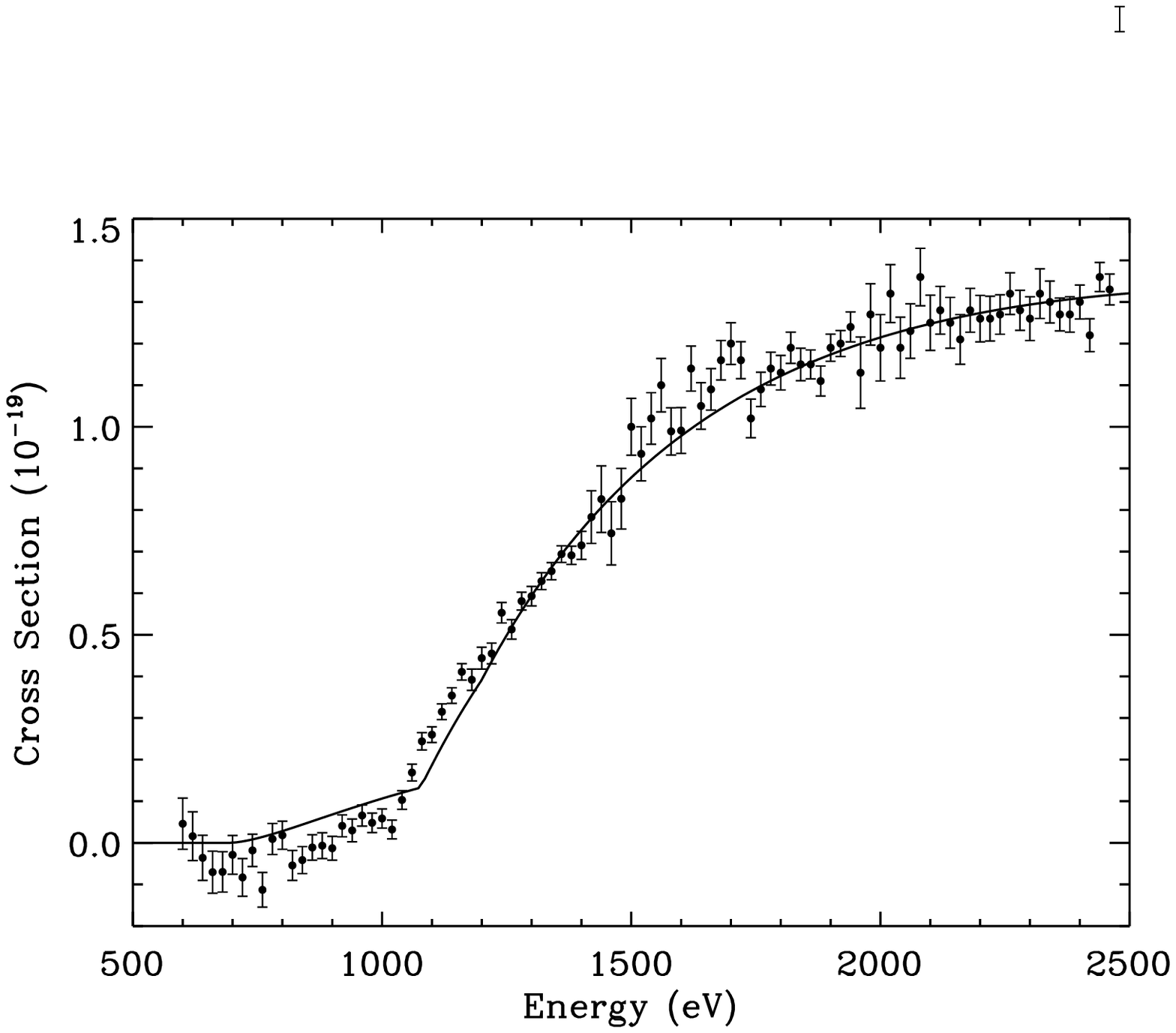}
\caption{\label{fig:fe1113} Double ionization of Fe$^{11+}$ forming Fe$^{13+}$. The data points show the experimental results of \citet{Hahn:ApJ:2011}. The solid curve is the cross section used here, which is described in Table~\ref{table:cross}.
}
\end{figure}

	Double-ionization measurements for $q_i=9, 11, 12$, and $13$ have been measured by \citet{Hahn:ApJ:2011,Hahn:ApJ:2011a,Hahn:ApJ:2012,Hahn:ApJ:2013}. These data show that there is a small contribution due to direct ionization, starting at the direct-double-ionization threshold. However, the dominant double-ionization process for highly charged Fe ions is initially single ionization of an $L$-shell ($n=2$) electron, followed by autoionization, resulting in a net double ionization. The relative importance of direct ionization compared to IA decreases as the charge state increases. We have found that a reasonable approximation to these experimental data is obtained by using Equation~(\ref{eq:shevtar}) to represent the direct contribution and including indirect ionization using the Lotz cross section for the $L$-shell ionization scaled by the branching ratios from \citet{Kaastra:AAS:1993}. An example of this semiempirical prediction compared to the experiment for the case of Fe$^{11+}$ forming Fe$^{13+}$ is shown in Figure~\ref{fig:fe1113}. We have extended this scheme beyond the several measured charge states to represent double ionization for Fe$^{7+}$--Fe$^{15+}$. For higher order EIMI, no experimental data exist so we use the Lotz formula scaled by the branching ratios. 
		
	For Fe$^{16+}$ and higher charge states, we are not aware of any experimental data for EIMI. For these ions, we estimate the double-ionization cross section using the semiempirical formulae of \citet{Shevelko:JPhysB:2005}, i.e., Equation~(\ref{eq:shev5a}) for direct ionization and Equation~(\ref{eq:shev5b}) for the $K$-shell IA. For these charge states, the branching ratios indicate the EIMI of higher order than double ionization is negligible. 

\section{Charge State Distribution}\label{sec:csd}

	The ion abundance $y_{i}$ of charge state $i$ as a function of time is described by
\begin{equation}
\frac{\mathrm{d}y_{i}}{\mathrm{d}t}=n_{\mathrm{e}}
\left[I^{j}_{i-j}y_{i-j}+\ldots+I^{1}_{i-1}y_{i-1}-\left(I^{1}_{i}+\ldots+I^{k}_{i}+R_{i}\right)y_{i}+R_{i+1}y_{i}\right], 
\label{eq:dydt}
\end{equation}
where $I^{j}_{i}$ is the rate coefficient for $j$-times ionization from charge state $i$ to $i+j$ and $R_{i}$ is the recombination rate coefficient from $i$ to $i-1$. The terms on the right represent, from left to right, ionization from lower charge states into $i$, ionization and recombination out of $i$ to other charge states, and recombination from $i+1$ into $i$. For most astrophysical plasmas, the density is low so that multiple recombination is extremely unlikely. Note also that in this expression, the rate coefficients $I$ and $R$ are functions of temperature. In a dynamic plasma, the temperature and density vary in time. 

	The rate coefficients for Equation~(\ref{eq:dydt}) were derived from several sources. The EISI rate coefficients come from the recommended data of \citet{Dere:AA:2007}. We have found these data to be in reasonable agreement with experiment for single ionization from Fe$^{7+}$ and Fe$^{9+}$--Fe$^{17+}$ \citep[][and references therein]{Hahn:JPCS:2014}. The radiative and dielectronic recombination rate coefficients are the ones compiled in the CHIANTI atomic database \citep{Dere:AAS:1997, Landi:ApJ:2013}. For iron, many of these recombination rate coefficients are based on the calculations of \citet{Badnell:AA:2003} and \citet{Badnell:ApJ:2006}\footnote{http://amdpp.phys.strath.ac.uk/tamoc/DR/ and http://amdpp.phys.strath.ac.uk/tamoc/RR/}. The dielectronic recombination data have been experimentally benchmarked by ion storage ring experiments \citep[][and references therein]{Schippers:IRAMP:2010}.
	
	The EIMI rate coefficients have been derived from the formulae given in Section~\ref{sec:cross} by numerically convolving the cross sections with a Maxwellian electron energy distribution. %The procedure has been described in \citet[e.g.,][]{Fogle:ApJS:2008}.
In dynamic or tenuous plasmas it is possible that the electron energy distribution is non-Maxwellian. Equilibrium-ionization-balance calculations for non-Maxwellian distributions have been given by \citet{Dzifcakova:ApJS:2013}, albeit without considering EIMI. Here, all of our calculations are based on Maxwellian distributions. 

\section{Equilibrium Ionization Balance}\label{sec:equilib}
	
	Collisional ionization equilibrium (CIE) occurs when the left hand side of Equation~(\ref{eq:dydt}) is zero. In this case the density is a constant factor and plays no role in the solution. For a given temperature, we have a system of algebraic equations. It is easy to see that Equation~(\ref{eq:dydt}) can be written as a matrix: 
\begin{equation}
\mathbf{A}\vec{y}=0, 
\label{eq:ay0}
\end{equation}
where $\mathbf{A}$ is the matrix of the rate coefficients and $\vec{y}$ is the vector of abundances with elements $y_{i}$. In order to obtain a unique solution, an additional equation is needed. For this, we require that abundances be normalized so that 
\begin{equation}
\sum_{i}{y_{i}}=1.
\label{eq:norm}
\end{equation}
This condition is implemented by replacing one of the rows of Equation~(\ref{eq:ay0}) with Equation~(\ref{eq:norm}), see for example \citet{Bryans:ApJS:2006}. 

\begin{figure}
\centering \includegraphics[width=0.9\textwidth]{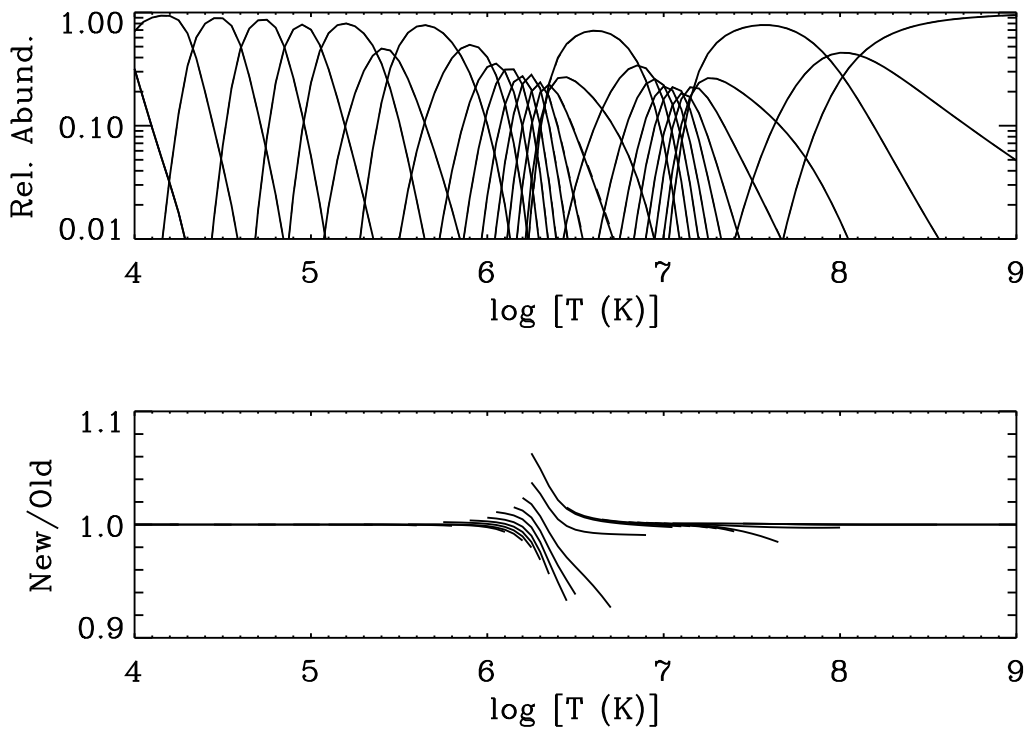}
\caption{\label{fig:equilib} The top panel shows the CIE charge balance for iron when multiple ionization is included (solid curves) and when it is ignored (dashed curves). The two cases are not distinguishable in this figure because the differences are very small. The bottom panel plots the ratio of the ion abundances from the new calculation, which includes EIMI, divided by the old calculations with only single ionization. EIMI changes the equilibrium abundances by 5\% at most. 
}
\end{figure}

	Figure~\ref{fig:equilib} shows the equilibrium ionization balance as a function of temperature, $y_{i}(T)$, for all the iron charge states. The solid lines in the figure show the results of the present calculation, which includes EIMI processes. The figure also includes a set of dashed curves, which show the results when only single ionization is considered, however the differences are smaller than the width of the lines in the figure. Here, our CIE calculations for single ionization are identical to the results given in the CHIANTI database, which are an updating of the CIE calculations of \citet{Bryans:ApJ:2009}. This agreement between our results and those of CHIANTI is expected since we used the same single ionization and recombination rate coefficients.
	
	The lower panel of Figure~\ref{fig:equilib} highlights the differences between CIE calculations including and excluding EIMI. The figure shows the ratio of the abundances calculated with EIMI divided by the abundances considering only EISI. The largest changes occur in the vicinity of Fe$^{16+}$, which is formed over a very broad temperature range. The effect of including EIMI versus EISI is to decrease the temperature at which Fe$^{16+}$ and the surrounding charge states are formed. The CSD for low charge states is not greatly affected by EIMI, because for these ions the ionization thresholds are spaced out such that multiple ionization becomes significant at temperatures where the abundance of that charge state has already become small due to EISI. Similarly, EIMI is less important for higher charge states because for those ions the direct ionization cross section is very small and they are open $L$-shell ions so that IA occurs mainly through the formation of $K$-shell holes, which requires ever higher energies relative to the EISI threshold. We find that the CIE calculations, including or excluding EIMI, agree to within 5\% for all the charge states. This demonstrates that, as expected, EIMI can be safely ignored in CIE calculations unless extremely high precision is required. 

\section{Dynamic Ionization Balance}\label{sec:dynamic}
	
	EIMI is expected to be more important in plasmas in which the electron temperature changes rapidly. As EIMI processes increase the ionization rate they can increase the rate at which the charge balance adjusts to sudden changes in the electron temperature.

\subsection{Equilibration Timescales}\label{subsec:timescales}

	One way to quantify the effect of EIMI processes on a dynamic plasma is to calculate the timescale for the charge balance to reach equilibrium following a sudden change in $T$. These timescales have been used, for example, in the analysis of spectra from supernova remnants \citep{Masai:ASS:1984, Hughes:ApJ:1985, Smith:ApJ:2010}. 
	
	The method for calculating the ionization timescales has been described by \citet{Masai:ASS:1984}. Equation~(\ref{eq:dydt}) can be written in the form
\begin{equation}
\frac{d\vec{y}}{dt}=n_{\mathrm{e}}\mathbf{A}(T)\vec{y}.
\label{eq:eqtime1}
\end{equation}
For constant $T$ and $n_{\mathrm{e}}$, the solution to this equation is $\vec{y}(t) = \vec{y}_{0}\exp{\left[n_{\mathrm{e}}t \mathbf{A}(T)\right]}$. The exponential of a matrix is defined in terms of a Taylor expansion, which involves powers of the matrix $\mathbf{A}$. This is simplified by diagonalizing the matrix by finding its eigenvalues and eigenvectors. Doing so, one finds that the solution to Equation~(\ref{eq:eqtime1}) can be written as 
\begin{equation}
\vec{y}(t)=\mathbf{S} \left[ \begin{array}{ccc}
\exp{\left(\lambda_{1}n_{\mathrm{e}}t\right)} & 0 & \cdots \\
0 & \exp{\left(\lambda_{j}n_{\mathrm{e}}t\right)} & 0 \cdots \\
\vdots & 0 & \ddots 
\end{array}\right]\mathbf{S}^{-1}\vec{y}_{0},
\label{eq:eqtime2}
\end{equation}
where $\mathbf{S}$ is the matrix in which each $j$th column is the eigenvector corresponding to the eigenvalue $\lambda_{j}$ of the diagonal matrix. Thus, the density-weighted timescales for equilibration are given by $1/\lambda_{j}$ in units of $n_{\mathrm{e}}t$. Note that the eigenvectors do not generally correspond to individual elements of $\vec{y}$ (i.e., charge states), but are instead linear combinations of those elements. 

\begin{figure}
\centering \includegraphics[width=0.9\textwidth]{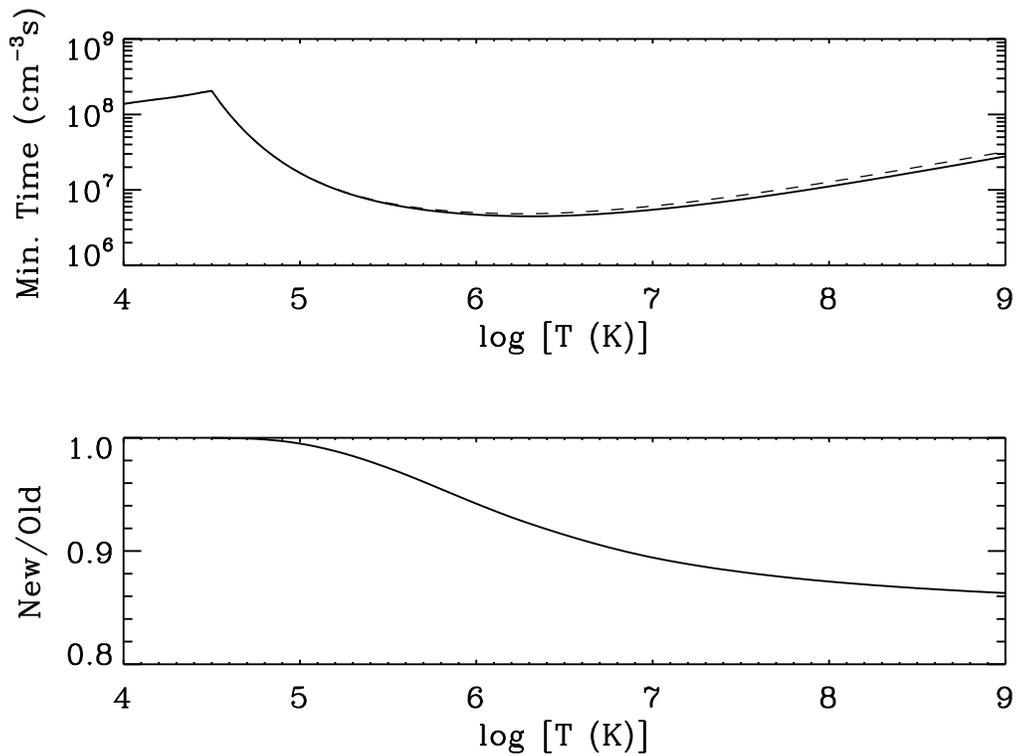}
\caption{\label{fig:mintime} The top panel shows the minimum scale $n_{\mathrm{e}}t$ for the ion balance to equilibrate following rapid heating to a given temperature $T$. The solid line represents the new calculations with EIMI included, while the dashed line indicates the results with only EISI. The bottom panel shows the ratio of the scales including (New) versus excluding (Old) EIMI.
}
\end{figure}

	Figure~\ref{fig:mintime} presents the results for the minimum $1/\lambda_{j}$ as a function of temperature. This represents the scale $n_{\mathrm{e}}t$ for any significant changes in the ion population to occur. The bottom panel of Figure~\ref{fig:mintime} presents the ratio of the scale when EIMI is included versus when it is ignored. These results show that EIMI causes the plasma to evolve faster at high temperatures. For example, for $T=10^{7}$~K the CSD begins to change about 10\% faster if EIMI is included in the calculation than when it is ignored. In contrast, the maximum $1/\lambda$ at a given temperature are nearly identical whether or not EIMI is considered in the calculation. Thus, EIMI causes changes to begin sooner, but the total time it takes for the system to asymptote to equilibrium is not significantly different when considering EIMI. 
	
	The reason the minimum $n_{\mathrm{e}}t$ is more sensitive to including EIMI channels compared to the maximum can be seen by looking at the eigenvectors. For a given $T$, the eigenvector corresponding to the minimum $n_{\mathrm{e}}t$ eigenvalue is a linear combination of the abundances from the lowest charge states. The eigenvector corresponding to the maximum $n_{\mathrm{e}}t$ is mainly made up of components from the charge states that are abundant in CIE at that temperature. Thus, EIMI influences the ion balance by significantly increasing the ionization rate from the lowest charge states. However, for charge states that are already close to equilibrium, the EISI cross sections are larger and so EIMI has little additional influence.
	
	Figure~\ref{fig:mintime} can be compared to Figure~2 of \citet{Smith:ApJ:2010}, where a similar calculation was carried out that included only single-ionization rates. There is a clear resemblance between the shape of the dependence of the scale on temperature. However, there are differences in magnitude. At low temperatures these differences are almost an order of magnitude, while at high temperatures the two calculations are nearly the same. These differences are probably due to the different ionization rate coefficients used in the calculations. \citet{Smith:ApJ:2010} used the ionization and recombination rate coefficients from \citet{Mazzotta:AAS:1998}, whereas we have used those from CHIANTI \citep{Dere:AA:2007,Landi:ApJ:2013}. \citet{Bryans:ApJ:2009} performed CIE calculations using the same single ionization rate coefficients that we use and compared the results to the CIE results derived from the \citet{Mazzotta:AAS:1998} rate coefficients. They also found order of magnitude differences at low temperatures when comparing these different data sources. This discrepancy demonstrates the importance of having reliable ionization and recombination rate coefficients.

\subsection{Direct Calculations}\label{subsec:numerical}

	For systems in which $T$ or $n_{\mathrm{e}}$ evolve in time, it is necessary to numerically solve Equation~(\ref{eq:dydt}) to find the CSD at each time step. One challenge in doing this is that Equation~(\ref{eq:dydt}) represents a ``stiff'' system of ordinary differential equations. That is, the coefficients on the right hand side of the equation can vary by orders of magnitude. Standard numerical methods have been developed for dealing with such stiff equations \citep[e.g.,][]{Press:Book}. In order to ensure accuracy, we have used an adaptive time step for the integration. Conditions for adapting the timestep have been given by \citet{MacNeice:SolPhys:1984} and used more recently by \citet{Bradshaw:AA:2009}. These conditions define values $\epsilon_d$ and $\epsilon_{r}$ and require that the time step be small enough that, for all $i$
\begin{equation}
|y_{i}(t+\Delta t)-y_{i}| \leq \epsilon_{d},
\label{eq:cond_epsilond}
\end{equation}
and 
\begin{equation}
|\log [y_{i}(t+\Delta t)] - \log [y_{i}(t)]| \leq \epsilon_{r}. 
\label{eq:cond_epsilonr}
\end{equation}
\citet{MacNeice:SolPhys:1984} and \citet{Bradshaw:AA:2009} have found that setting $\epsilon_r = 0.6$ and $\epsilon_d=0.1$ are good control parameters. One check on the accuracy of the numerical solution is that $\sum_{i} y_{i} = 1$. We find in our results that the total abundance differs from unity by less than one part in $10^{11}$, where we have not imposed any additional normalization at each time step. 

\begin{figure}
\centering \includegraphics[width=0.9\textwidth]{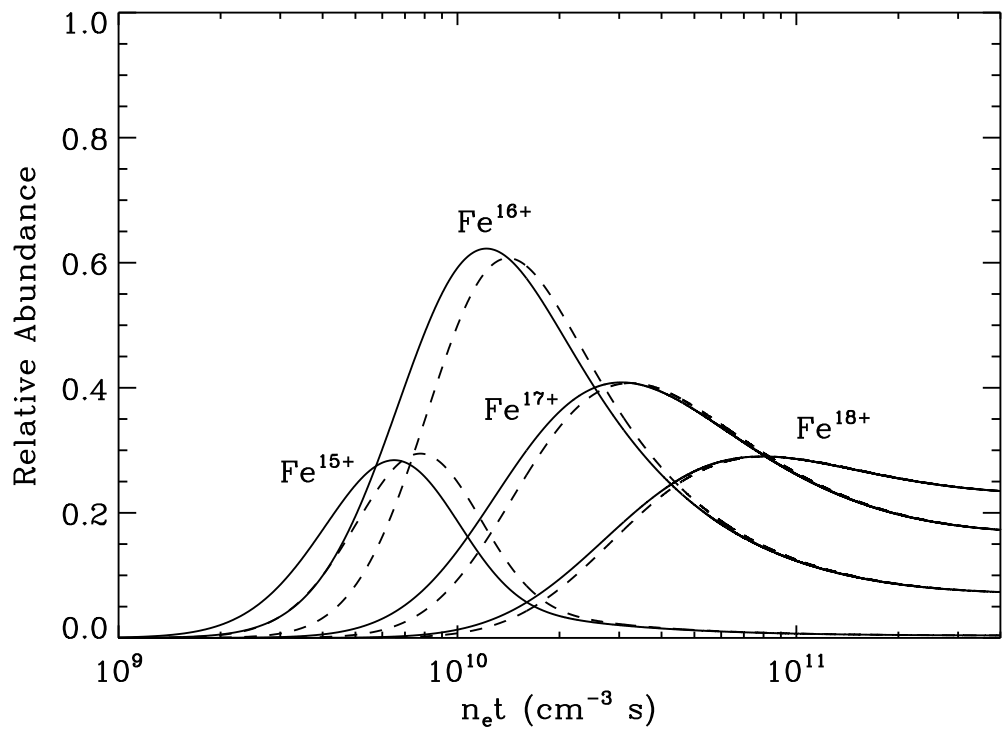}
\caption{\label{fig:jump} Ion abundances of Fe$^{15+}$ to Fe$^{18+}$ versus scale $n_{\mathrm{e}}t$ following a sudden jump in temperature from $10^{5}$~K to $10^{7}$~K. The solid curves indicate the results with EIMI and the dashed curves are the calculations including only EISI. It is clear that EIMI causes the charge states to evolve faster than if it is ignored.
}
\end{figure}

	Figure~\ref{fig:jump} shows the time evolution of the abundances of selected charge states following a sudden jump in $T$ from $10^{5}$~K to $10^{7}$~K at a density of $n_{\mathrm{e}}=10^{9}$~$\mathrm{cm^{-3}}$. This is essentially the same scenario as was described above in Section~\ref{subsec:timescales}, but using a direct calculation clarifies the relation between the timescales and charge state abundances. These results show that EIMI allows the CSD to evolve more rapidly than if only EISI is considered. Nevertheless, this change is relatively small, being faster in this case by only a few percent. At large $n_{\mathrm{e}}t$, the abundances asymptote to their CIE values. 
	
\subsection{Application to Nanoflares}\label{subsec:nanoflares}	
	
	A much different situation can arise when the temperature is oscillating. In this case, the system can be prevented from reaching an equilibrium and the effects of EIMI are more important. To illustrate these effects for a particular case in astrophysics, we consider some parameters that are relevant for nanoflare heating of the solar corona. 
	
	One theory for the heating of the solar corona is that it is caused by numerous relatively small impulsive heating events known as nanoflares. These are usually thought to be caused by magnetic reconnection, although other processes could have a similar impulsive character, such as resonant wave absorption \citep{Klimchuk:SolPhys:2006}. Recently, there has been significant work on predicting the spectroscopic signatures of nanoflares \citep[e.g.,][]{Bradshaw:AA:2003,Cargill:ApJ:2004,Reale:ApJ:2008, Bradshaw:ApJS:2011}. Nanoflares are predicted to heat the plasma to $\sim10^{7}$~K \citep{Schmelz:ApJ:2009,Brosius:ApJ:2014}. However, simulations predict such hot plasma difficult to detect because the ionization balance needs time to adjust to the high temperature \citep{Bradshaw:ApJS:2011}. More detailed models have attempted to predict the temperature distribution of nanoflare heated plasma \citep{Bradshaw:ApJ:2012, Reep:ApJ:2013}. All of these computations have so far neglected EIMI. 
	
\begin{figure}
\centering \includegraphics[width=0.9\textwidth]{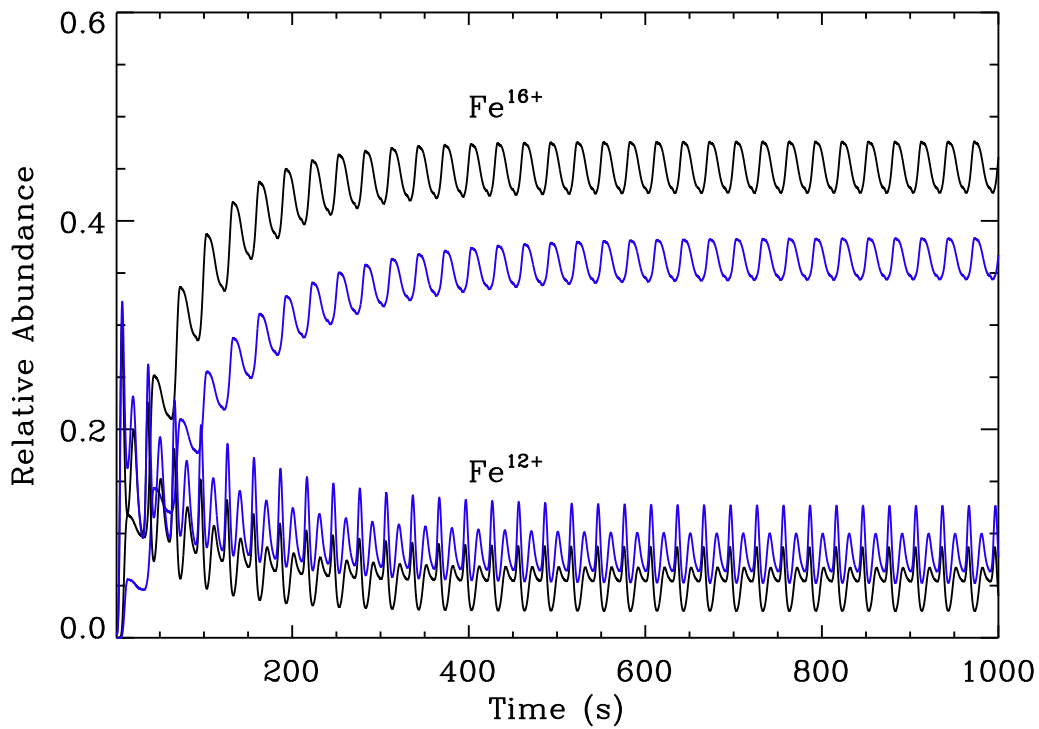}
\caption{\label{fig:sine} Ion abundances of Fe$^{12+}$ and Fe$^{16+}$ for a temperature oscillating between $10^{5}$ and $10^{7}$~K with a period of 20~s at $n_{\mathrm{e}}=5 \times 10^{8}$~cm$^{-3}$. The black curve indicates the results that include EIMI, while the blue curve includes only EISI. 
}
\end{figure}

	In order to estimate the possible effects of including EIMI in nanoflare models, we have performed dynamic ionization balance calculations with an oscillating temperature. Figure~\ref{fig:sine} shows one example of our results. Here the temperature is oscillating so that $\log T[\mathrm{K}] = 6.0 \sin(2\pi t/\tau)$, where $\tau=20$~s is the period of the oscillation. This period is comparable to that used in other nanoflare calculations. For example, \citet{Reep:ApJ:2013} consider a series of nanoflares with heating timescales of 60~s. \citet{Klimchuk:ApJ:2014} have considered even faster timescales, with 10~s duration heating events. The temperature in our calculation varies between 10$^{5}$ and 10$^{7}$~K, which are reasonable values for the solar transition region and corona. For this calculation we have set $n_{\mathrm{e}}=5 \times 10^{8}$~cm$^{-3}$, which is a typical density for the low solar corona. It is worth noting, however, that we have ignored essentially all of the hydrodynamics involved in simulating nanoflares. In reality, the heating will drive corresponding changes in the plasma density that will in turn modify the temperature. Our objective here is only to determine whether or not the neglect of EIMI can be justified. 
	
	Figure~\ref{fig:sine} shows that the oscillating temperature prevents the CSD from asymptoting to the CIE values. Instead, after a short transient at the beginning of the simulation, the CSD settles down into a stable oscillation around an average abundance value. 
	
\begin{figure}
\centering \includegraphics[width=0.9\textwidth]{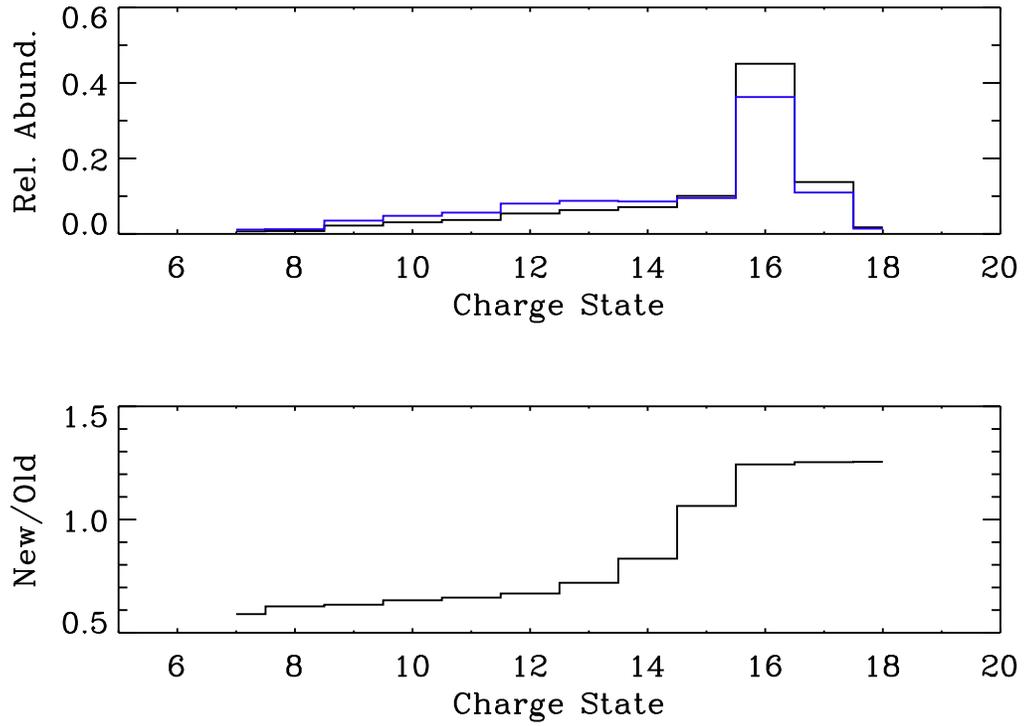}
\caption{\label{fig:sineabund} The average CSD for time $t > 1000$~s, for the same oscillating temperatures as for Figure~\ref{fig:sine}. The black curve shows the results with EIMI while the blue curve includes only EISI. The average abundances differ by up to 40\%.
}
\end{figure}

	These average abundances can be significantly different depending on whether EIMI is included or neglected. Figure~\ref{fig:sineabund} shows the average abundances of the iron charge states for times $t > 1000$~s, when the oscillation is stable. This reveals that neglecting EIMI overestimates the abundances of charge states below Fe$^{15+}$ and underestimates the abundance of Fe$^{15+}$ and higher charge states. These differences are up to 40\%, for those ions having significant relative abundances ($y_{i}>0.01$). The size of the discrepancy depends on the timescale of the oscillation. With a shorter period of about $\tau=10$~s, the discrepancy is about 50\%, while for a longer period of $\tau=60$~s, the difference is about 20\%. 
	
	There are clear implications for spectroscopic diagnostics searching for nanoflares. Models that neglect EIMI will systematically underestimate the abundances of charge states Fe$^{15+}$ and above. This means that observations will see more high temperature plasma than is currently predicted. Additionally, differential emission measure spectroscopic analyses will have a higher ratio of hot to warm plasma than is predicted by current models. Thus, our results suggests that EIMI should be considered in nanoflare model calculations in order to accurately predict observed spectra. 
	
\section{Conclusions}\label{sec:conclusions}

	It has been thought that EIMI would be unimportant in many astrophysical contexts, especially in situations close to CIE. Combined with the lack of EIMI cross section data, multiple-ionization processes have been generally ignored. In order to determine whether EIMI can or cannot be neglected, we have studied iron charge-state abundances by incorporating EIMI cross sections into calculations for CIE, equilibration timescales, and time-dependent ionization. 
	
	We find that for CIE it is justified to ignore EIMI, as the influence is less than 5\% for iron. EIMI has a more significant influence on the CSD of ionizing plasmas and can decrease the timescale at which changes in the CSD begin to occur by $\sim 10\%$ at temperatures around 10$^{7}$~K. However, currently the uncertainties in EISI cross sections are likely to be more important than whether EIMI is included or neglected. 
	
	The greatest change we found when including EIMI is for an oscillation in the temperature. In this scenario we found cases where the ion abundances may differ by up to 50\% from what is predicted when EIMI is neglected. One context in which such temperature oscillations occur is nanoflare heating of the solar corona. Based on our results, nanoflare models should incorporate EIMI in order to accurately predict the spectrosopic signatures of nanoflares. 
	
	A challenge for incorporating EIMI into plasma models is the lack of any reliable EIMI theory and the dearth of experimental measurements. Here we have focused on iron, for which at least double-ionization measurements exist for enough ions to reasonably interpolate the cross sections using semiempirical formulae. For most other ions the situation is worse. Nevertheless, we can speculate about the influence of EIMI on the CSD for other elements. We find significant EIMI is due mainly to direct ionization or $L$-shell IA, while $K$-shell IA occurs at such high energies that it has little effect. This suggests that the CSD of elements below Na, which have open $L$-shells, will be less sensitive to EIMI. In contrast, EIMI may become more important for elements heavier than iron as additional EIMI channels become possible. 
	
\begin{acknowledgments}
This work was supported in part by the NASA Solar Heliospheric Physics program grant NNX09AB25G and the NSF Division of Atmospheric and Geospace Sciences SHINE program grant AGS-1060194. 
\end{acknowledgments}

\begin{deluxetable}{l c c c c c c c c c c}
\tabletypesize{\small}
\tablecolumns{11}
\tablewidth{0pc}
\tablecaption{Fitting Formulae for Iron Ionization Cross Sections
\label{table:cross}}
\tablehead{
\colhead{$q_{i}$} &
\colhead{$q_{f}$} &
\colhead{Process} & 
\colhead{Equation} & 
\colhead{$E_{\mathrm{th}}$} & 
\colhead{$f$} &
\colhead{$p_{0}$} &
\colhead{$p_{1}$} &
\colhead{$p_{2}$} &
\colhead{$p_{3}$} & 
\colhead{Expt. Reference} 
}
\startdata
0	& 2 & D		& \ref{eq:shev6a} 	& 24.10	& 1.0			& 11.0119 &				& 			& & \citet{Shah:JPhysB:1993} \\
	& 	& IA 	& \ref{eq:shev6b} 	& 59.0	& 1.0			& 3.20615 & 3.0		&				&	& \ditto \\
	& 	& IA 	& \ref{eq:lotz}			& 7117.0& 0.1005	& 2.0			& 			& 			& & \nodata \\
0 & 3 & D		& \ref{eq:shevtar} 	& 54.75 & 0.0860	& 6.30		& 26.0	& 1.20	& 1.0 & \citet{Shah:JPhysB:1993} \\
	& 	& IA	& \ref{eq:lotz}			& 98.0	& 0.9359	& 2.0			&				& 			& & \ditto \\
	& 	& IA 	& \ref{eq:lotz}			& 713.0	& 0.3086	& 4.0			&				& 			& & \ditto \\
	& 	& IA 	& \ref{eq:lotz}			& 726.0	& 0.3096	& 2.0			&				& 			& & \ditto \\
	& 	& IA	& \ref{eq:lotz}			& 851.0	& 0.0893	& 2.0			&				& 			& & \ditto \\
	& 	& IA	& \ref{eq:lotz}			& 7117.0& 0.0984	& 2.0			&				& 			& & \nodata \\
0	& 4	& D		& \ref{eq:shevtar} 	& 109.66& 1.0			& 0.5			& 26.0	& 1.73	& 1.0 & \citet{Shah:JPhysB:1993} \\
0	& 5	& IA	& \ref{eq:lotz}			& 713.0	& 0.0966	& 4.0			& 			& 			& & \nodata \\
	& 	& IA	& \ref{eq:lotz}			& 726.0	& 0.0802	& 2.0			& 			& 			& & \nodata \\
	& 	& IA	& \ref{eq:lotz}			& 851.0	& 0.3317	& 2.0			& 			& 			& & \nodata \\
	& 	& IA	& \ref{eq:lotz}			& 7117.0& 0.0970	& 2.0			& 			&				& & \nodata \\
0	& 6	& IA	& \ref{eq:lotz}			& 713.0	& 0.0047	& 4.0			& 			& 			& & \nodata \\
	& 	& IA	& \ref{eq:lotz}			& 726.0	& 0.0039	& 2.0			& 			& 			& & \nodata \\
	& 	& IA	& \ref{eq:lotz}			& 851.0	& 0.2495	& 2.0			& 			& 			& & \nodata \\
	& 	& IA	& \ref{eq:lotz}			& 7117.0& 0.1598	& 2.0			& 			& 			& & \nodata \\
0	& 7 & IA	& \ref{eq:lotz}			& 851.0	& 0.1122	& 2.0			& 			& 			& & \nodata \\
	& 	& IA	& \ref{eq:lotz}			& 7117.0& 0.2007	& 2.0			& 			& 			& & \nodata \\
0	& 8	& IA	& \ref{eq:lotz}			& 7117.0& 0.1729	& 2.0			& 			& 			& & \nodata \\
0	& 9	& IA	& \ref{eq:lotz}			& 7117.0& 0.0205	& 2.0			& 			& 			& & \nodata \\
0 & 10& IA	& \ref{eq:lotz}			& 7117.0& 0.0022	& 2.0			& 			& 			& & \nodata \\
1	& 3 & D		& \ref{eq:shev6a}		& 43.54 & 1.0			& 17.0		& 			& 			& & \citet{Stenke:JPhysB:1999} \\
	& 	& IA	& \ref{eq:shev6b}		& 69.58	& 1.0			& 3.72		& 3.0		& 			& & \ditto \\	
	& 	& IA	& \ref{eq:shev6b}		& 112.85& 1.0			& 3.31		& 3.0		& 			& & \ditto \\
	& 	& IA	& \ref{eq:shev6b}		& 727.64& 1.0			& 1.0			& 2.0		& 			& & \ditto \\
	& 	& IA	& \ref{eq:shev6b}		& 868.24& 1.0			& 1.0			& 2.0		& 			& & \ditto \\
	& 	& IA	& \ref{eq:lotz}			& 7164.0& 0.1007	& 2.0			& 			& 			& & \nodata \\
1	& 4	& IA	& \ref{eq:lotz}			& 119.0	& 0.9563 	& 2.0			& 			& 			& & \nodata \\
	& 	& IA	& \ref{eq:lotz}			& 744.0	& 0.3061	& 4.0			& 			& 			& & \nodata \\
	& 	& IA	& \ref{eq:lotz}			& 757.0	& 0.3076	& 2.0			& 			& 			& & \nodata \\
	& 	& IA	& \ref{eq:lotz}			& 882.0	& 0.0953	& 2.0			& 			& 			& & \nodata \\
	& 	& IA	& \ref{eq:lotz}			& 7164.0& 0.0976	& 2.0			& 			& 			& & \nodata \\
1	& 5	& IA	& \ref{eq:lotz} 		& 744.0	& 0.4256	& 4.0			& 			& 			& & \nodata \\
	& 	& IA	& \ref{eq:lotz} 		& 757.0	& 0.3542	& 2.0			& 			& 			& & \nodata \\
	& 	& IA	& \ref{eq:lotz}			& 882.0	& 0.2581	& 2.0			& 			& 			& & \nodata \\
	& 	& IA	& \ref{eq:lotz}			& 7164.0& 0.1564	& 2.0			& 			& 			& & \nodata \\
1 & 6	& IA	& \ref{eq:lotz}			& 744.0	& 0.0759	& 4.0			& 			& 			& & \nodata \\
	& 	& IA	& \ref{eq:lotz}			& 757.0	& 0.0617	& 2.0			& 			& 			& & \nodata \\
	& 	& IA	& \ref{eq:lotz}			& 882.0	& 0.4829	& 2.0			& 			& 			& & \nodata \\
	& 	& IA	& \ref{eq:lotz}			& 7164.0& 0.3242	& 2.0			& 			& 			& & \nodata \\
1	& 7 & IA	& \ref{eq:lotz}			& 744.0	& 0.0001	& 4.0			& 			& 			& & \nodata \\
	& 	& IA	& \ref{eq:lotz}			& 757.0	& 0.0001	& 2.0			& 			& 			& & \nodata \\
	& 	& IA	& \ref{eq:lotz}			& 882.0	& 0.1409	& 2.0			& 			& 			& & \nodata \\
	& 	& IA	& \ref{eq:lotz}			& 7164.0& 0.0.3242& 2.0			& 			& 			& & \nodata \\
1	& 8	& IA	& \ref{eq:lotz}			& 882.0	& 0.0058	& 2.0			& 			& 			& & \nodata \\
	& 	& IA	& \ref{eq:lotz}			& 7164.0& 0.1154	& 2.0			& 			& 			& & \nodata \\
1	& 9	& IA	& \ref{eq:lotz}			& 7164.0& 0.0143	& 2.0			& 			& 			& & \nodata \\
1 & 10& IA	& \ref{eq:lotz}			& 7164.0& 0.0007	& 2.0			& 			& 			& & \nodata \\
2	& 4	& D		& \ref{eq:shevtar}	& 85.56	& 1.0			& 14.0		& 24.0	& 1.08	& 0.75 & \nodata \\
	& 	& IA	& \ref{eq:lotz}			& 141.0	& 0.0217	& 2.0			& 			& 			& & \nodata \\
	& 	& IA	& \ref{eq:lotz}			& 775.0	& 0.1773	& 4.0			& 			& 			& & \nodata \\
	& 	& IA	& \ref{eq:lotz}			& 787.9	& 0.2612	& 2.0			& 			& 			& & \nodata \\
	& 	& IA	& \ref{eq:lotz}			& 913.0	& 0.0194	& 2.0			& 			& 			& & \nodata \\
	& 	& IA	& \ref{eq:lotz}			& 7210.0& 0.1009	& 2.0			& 			& 			& & \nodata \\
2	& 5	& IA	& \ref{eq:lotz}			& 775.0	& 0.7056	& 4.0			& 			& 			& & \nodata \\
	& 	& IA	& \ref{eq:lotz}			& 787.9	& 0.5912	& 2.0			& 			& 			& & \nodata \\
	& 	& IA	& \ref{eq:lotz}			& 913.0	& 0.3687	& 2.0			& 			& 			& & \nodata \\
	& 	& IA	& \ref{eq:lotz}			& 7210.0& 0.2220	& 2.0			& 			& 			& & \nodata \\	
2 & 6	& IA	& \ref{eq:lotz}			& 775.0	& 0.1018	& 4.0			& 			& 			& & \nodata \\
	& 	& IA	& \ref{eq:lotz}			& 787.9	& 0.1317	& 2.0			& 			& 			& & \nodata \\
	& 	& IA	& \ref{eq:lotz}			& 913.0	& 0.4976	& 2.0			& 			& 			& & \nodata \\
	& 	& IA	& \ref{eq:lotz}			& 7210.0&	0.2139	& 2.0			& 			&				& & \nodata \\
2	& 7	& IA	& \ref{eq:lotz}			& 913.0	& 0.1139	& 2.0			& 			& 			& & \nodata \\
	& 	& IA	& \ref{eq:lotz} 		& 7210.0& 0.3601	& 2.0			& 			& 			& & \nodata \\
2 & 8 & IA	& \ref{eq:lotz}			& 913.0	& 0.0004	& 2.0			& 			& 			& & \nodata \\
	& 	& IA	& \ref{eq:lotz}			& 7210.0& 0.0858	& 2.0			& 			& 			& & \nodata \\
2	& 9	& IA	& \ref{eq:lotz}			& 7210.0& 0.0123	& 2.0			& 			& 			& & \nodata \\
2	& 10& IA	& \ref{eq:lotz}			& 7210.0& 0.0003	& 2.0			& 			& 			& & \nodata \\	
3	& 5	& D		& \ref{eq:shev6a}		& 129.8	& 1.0			& 359.0		& 			& 			& & \citet{Stenke:JPhysB:1999} \\
	& 	& IA	& \ref{eq:shev6b}		& 142.1	& 1.0			& 1.0			& 3.0		& 			& & \ditto \\
	& 	& IA	& \ref{eq:shev6b}		& 760.7	& 1.0			& 1.0			& 2.0		& 			& & \ditto \\
	& 	& IA	& \ref{eq:shev6b}		& 900.5	& 1.0			& 1.0			& 2.0		& 			& & \ditto \\
	& 	& IA	& \ref{eq:lotz}			& 7256.0& 0.3058	& 2.0			& 			& 			& & \nodata \\
3	& 6	& IA	& \ref{eq:lotz} 		& 807.0 & 0.1333	& 4.0			& 			& 			& & \nodata \\
	& 	& IA	& \ref{eq:lotz}			& 819.9	& 0.1045	& 2.0			& 			& 			& & \nodata \\
	& 	& IA	& \ref{eq:lotz}			& 943.0	& 0.8295	& 2.0			& 			& 			& & \nodata \\
	& 	& IA	& \ref{eq:lotz}			& 7256.0& 0.1243	& 2.0			& 			& 			& & \nodata \\
3	& 7 & IA	& \ref{eq:lotz}			& 807.0	& 0.0067	& 4.0			& 			& 			& & \nodata \\
	& 	& IA	& \ref{eq:lotz}			& 819.9	& 0.0054	& 2.0			& 			& 			& & \nodata \\
	& 	& IA	& \ref{eq:lotz}			& 943.0	& 0.1426	& 2.0			& 			& 			& & \nodata \\
	& 	& IA	& \ref{eq:lotz}			& 7256.0& 0.3313	& 2.0			& 			& 			& & \nodata \\
3	& 8	& IA	& \ref{eq:lotz}			& 943.0	& 0.0004	& 2.0			& 			& 			& & \nodata \\
	& 	& IA	& \ref{eq:lotz}			& 7256.0& 0.1850	& 2.0			& 			& 			& & \nodata \\
3	& 9	& IA	& \ref{eq:lotz}			& 7256.0& 0.0398	& 2.0			& 			& 			& & \nodata \\
3 & 10& IA	& \ref{eq:lotz}			& 7256.0& 0.0097	& 2.0			& 			& 			& & \nodata \\
4	& 6	& D		& \ref{eq:shev6a}		& 174.1	& 1.0			& 303.0		& 			& 			& & \citet{Stenke:JPhysB:1999} \\
	& 	& IA	& \ref{eq:shev6b}		& 786.2	& 1.0			& 0.1			& 2.0		& 			& & \ditto \\
	& 	& IA	& \ref{eq:shev6b}		& 927.1	& 1.0			& 0.01		& 2.0		& 			& & \ditto \\
	& 	& IA	& \ref{eq:lotz}			& 7301.0& 0.0415	& 2.0			& 			& 			& & \nodata \\
4	& 7 & IA	& \ref{eq:lotz}			& 839.0	& 0.1633	& 4.0			& 			& 			& & \nodata \\
	& 	& IA	& \ref{eq:lotz}			& 851.9	& 0.1230	& 2.0			& 			& 			& & \nodata \\
	& 	& IA	& \ref{eq:lotz}			& 973.0	& 0.9741	& 2.0			& 			& 			& & \nodata \\
	& 	& IA	& \ref{eq:lotz}			& 7301.0& 0.1625	& 2.0			& 			& 			& & \nodata \\
4	& 8	& IA	& \ref{eq:lotz}			& 7301.0& 0.3961	& 2.0			& 			& 			& & \nodata \\
4	& 9	& IA	& \ref{eq:lotz}			& 7301.0& 0.1180	& 2.0			& 			& 			& & \nodata \\
4	& 10& IA	& \ref{eq:lotz}			& 7301.0& 0.0195	& 2.0			& 			& 			& & \nodata \\
5	& 7	& D		& \ref{eq:shev6a}		& 222.3	& 1.0			& 281.5		& 			& 			& &  \citet{Stenke:JPhysB:1999} \\
	& 	& IA	& \ref{eq:shev6b}		& 815.9	& 1.0			& 2.0			& 2.0		& 			& & \ditto \\
	& 	& IA	& \ref{eq:shev6b}		& 956.4	& 1.0			& 11.0		& 2.0		& 			& & \ditto \\
	& 	& IA	& \ref{eq:lotz}			& 7348.0& 0.3117	& 2.0			& 			& 			& & \nodata \\
5	& 8	& IA	& \ref{eq:lotz}			& 1003.0& 0.8690	& 2.0			& 			& 			& & \nodata \\
	& 	& IA	& \ref{eq:lotz}			& 7348.0& 0.1167	& 2.0			& 			& 			& & \nodata \\
5	& 9	& IA	& \ref{eq:lotz}			& 7348.0& 0.4220	& 2.0			& 			& 			& & \nodata \\
5 & 10& IA	& \ref{eq:lotz}			& 7348.0& 0.1085	& 2.0			& 			& 			& & \nodata \\
6	& 8	& D		& \ref{eq:shev6a}		& 286.0	& 1.0			& 261.3		& 			& 			& & \citet{Stenke:JPhysB:1999} \\
	& 	& IA	& \ref{eq:shev6b}		& 847.2	& 1.0			& 8.0			& 2.0		& 			& & \ditto \\
	& 	& IA	& \ref{eq:shev6b}		& 988.6	& 1.0			& 2.0			& 2.0		& 			& & \ditto \\
	& 	& IA	& \ref{eq:lotz}			& 7394.0& 0.3143	& 2.0			& 			& 			& & \nodata \\
6	& 9	& IA	& \ref{eq:lotz}			& 1033.0& 0.7790	& 2.0			& 			& 			& & \nodata \\
	& 	& IA	& \ref{eq:lotz}			& 7394.0& 0.1492	& 2.0			& 			& 			& & \nodata \\
6	& 10& IA	& \ref{eq:lotz}			& 7394.0& 0.4956	& 2.0			& 			& 			& & \nodata \\
7	& 9	& D		& \ref{eq:shevtar}	& 384.66& 1.0			& 14.0		& 19.0	& 1.08	& 0.75 & \nodata \\
	& 	& IA	& \ref{eq:lotz}			& 933.0	& 0.9967	& 4.0			& 			& 			& & \nodata \\
	& 	& IA	& \ref{eq:lotz}			& 945.8	& 0.9901	& 2.0			& 			& 			& & \nodata \\
	& 	& IA	& \ref{eq:lotz}			& 1094.0& 0.9187	& 2.0			& 			& 			& & \nodata \\
	& 	& IA	& \ref{eq:lotz}			& 7440.0& 0.3153	& 2.0			& 			& 			& & \nodata \\
7	& 10& IA	& \ref{eq:lotz}			& 7440.0& 0.1487	& 2.0			& 			& 			& & \nodata \\
7 & 11& IA	& \ref{eq:lotz}			& 7440.0& 0.4959	& 2.0			& 			& 			& & \nodata \\
8	& 10& D		& \ref{eq:shevtar}	& 495.70& 1.0			& 14.0		& 18.0	& 1.08	& 0.75 & \nodata \\
	& 	& IA	& \ref{eq:lotz}			& 965.0	& 0.9983	& 4.0			& 			& 			& & \nodata \\
	& 	& IA	& \ref{eq:lotz}			& 977.8	& 0.9983	& 2.0			& 			& 			& & \nodata \\
	& 	& IA	& \ref{eq:lotz}			& 1094.0& 0.9105	& 2.0			& 			& 			& & \nodata \\
	& 	& IA	& \ref{eq:lotz}			& 7486.0& 0.3165	& 2.0			& 			& 			& & \nodata \\
8	& 11& IA	& \ref{eq:lotz}			& 7486.0& 0.1471	& 2.0			& 			& 			& & \nodata \\
8 & 12& IA	& \ref{eq:lotz}			& 7486.0& 0.4975	& 2.0			& 			& 			& & \nodata \\
9	& 11& D		& \ref{eq:shevtar}	& 553.00& 1.0			& 14.0		& 17.0	& 1.08	& 0.75 &  \citet{Hahn:ApJ:2012} \\
	& 	& IA	& \ref{eq:lotz}			& 1000.0& 0.9973	& 4.0			& 			& 			& & \ditto \\
	& 	& IA	& \ref{eq:lotz}			& 1013.0& 0.9980	& 2.0			& 			& 			& & \ditto \\
	& 	& IA	& \ref{eq:lotz}			& 1129.0& 0.9214	& 2.0			& 			& 			& & \ditto \\
	& 	& IA	& \ref{eq:lotz}			& 7535.0& 0.3222	& 2.0			& 			& 			& & \nodata \\
9 & 12& IA	& \ref{eq:lotz}			& 7535.0& 0.1357	& 2.0			& 			& 			& & \nodata \\
9	& 13& IA	& \ref{eq:lotz}			& 7535.0& 0.5085	& 2.0			& 			& 			& & \nodata \\
10& 12& D		& \ref{eq:shevtar}	& 621.70& 1.0			& 14.0		& 16.0	& 1.08	& 0.75 & \nodata \\
	& 	& IA	& \ref{eq:lotz}			& 1036.0& 0.9952	& 4.0			& 			& 			& & \nodata \\
	& 	& IA	& \ref{eq:lotz}			& 1049.0& 0.9975	& 2.0			& 			& 			& & \nodata \\
	& 	& IA	& \ref{eq:lotz}			& 1164.0& 0.9154	& 2.0			& 			& 			& & \nodata \\
	& 	& IA	& \ref{eq:lotz}			& 7585.0& 0.3283	& 2.0			& 			& 			& & \nodata \\
10& 13& IA	& \ref{eq:lotz}			& 7585.0& 0.1267	& 2.0			& 			& 			& & \nodata \\
10& 14& IA	& \ref{eq:lotz}			& 7585.0& 0.5168	& 2.0			& 			& 			& & \nodata \\
11& 13& D		& \ref{eq:shevtar}	& 691.80& 1.0			& 14.0		& 15.0	& 1.08	& 0.75 & \citet{Hahn:ApJ:2011} \\
	& 	& IA	& \ref{eq:lotz}			& 1073.0& 0.9888	& 4.0			& 			& 			& & \ditto \\
	& 	& IA	& \ref{eq:lotz}			& 1086.0& 0.9967	& 2.0			& 			& 			& & \ditto \\
	& 	& IA	& \ref{eq:lotz}			& 1199.0& 0.9202	& 2.0			& 			& 			& & \ditto \\
	& 	& IA	& \ref{eq:lotz}			& 7636.0& 0.3345	& 2.0			& 			& 			& & \nodata \\
11& 14& IA	& \ref{eq:lotz}			& 7636.0& 0.1299	& 2.0			& 			& 			& & \nodata \\
11& 15& IA	& \ref{eq:lotz}			& 7636.0& 0.5120	& 2.0			& 			& 			& & \nodata \\
12& 14& D		& \ref{eq:shevtar}	& 753.20& 1.0			& 14.0		& 14.0	& 1.08	& 0.75 & \citet{Hahn:ApJ:2011a} \\
	& 	& IA	& \ref{eq:lotz}			& 1110.0& 0.9338	& 4.0			& 			& 			& & \ditto \\
	& 	& IA	& \ref{eq:lotz}			& 1123.0& 0.9954	& 2.0			& 			& 			& & \ditto \\
	& 	& IA	& \ref{eq:lotz}			& 1234.0& 0.9286	& 2.0			& 			& 			& & \ditto \\
	& 	& IA	& \ref{eq:lotz}			& 7686.0& 0.3336	& 2.0			& 			& 			& & \nodata \\
12& 15& IA	& \ref{eq:lotz}			& 7686.0& 0.1792	& 2.0			& 			& 			& & \nodata \\
12& 16& IA	& \ref{eq:lotz}			& 7686.0& 0.4575	& 2.0			& 			& 			& & \nodata \\
13& 15& D		& \ref{eq:shevtar}	& 848.40& 1.0			& 14.0		& 13.0	& 1.08	& 0.75 & \citet{Hahn:ApJ:2013} \\
	& 	& IA	& \ref{eq:lotz}			& 1147.0& 0.9046	& 4.0			& 			& 			& & \ditto \\
	& 	& IA	& \ref{eq:lotz}			& 1160.0& 0.9843	& 2.0			& 			& 			& & \ditto \\
	& 	& IA	& \ref{eq:lotz}			& 1270.0& 0.9452	& 2.0			& 			& 			& & \ditto \\
	& 	& IA	& \ref{eq:lotz}			& 7737.0& 0.3390	& 2.0			& 			& 			& & \nodata \\
13& 16& IA	& \ref{eq:lotz}			& 7737.0& 0.6297	& 2.0			& 			& 			& & \nodata \\
14& 16& D		& \ref{eq:shevtar}	& 945.51& 1.0			& 14.0		& 12.0	& 1.08	& 0.75 & \nodata \\
	& 	& IA	& \ref{eq:lotz}			& 1185.0& 0.8737	& 4.0			& 			& 			& & \nodata \\
	& 	& IA	& \ref{eq:lotz}			& 1198.0& 0.8737	& 2.0			& 			& 			& & \nodata \\
	& 	& IA	& \ref{eq:lotz}			& 1307.0& 0.9981	& 2.0			& 			& 			& & \nodata \\
	& 	& IA	& \ref{eq:lotz}			& 7788.0& 0.3985	& 2.0			& 			& 			& & \nodata \\
14& 17& IA	& \ref{eq:lotz}			& 7788.0& 0.5562	& 2.0			& 			& 			& & \nodata \\
15& 17& D		& \ref{eq:shevtar}	& 1752.0& 1.0			& 14.0		& 11.0	& 1.08	& 0.75 & \nodata \\
	& 	& IA	& \ref{eq:lotz}			& 7838.0& 0.6362	& 2.0			& 			& 			& & \nodata \\
16& 18& D		& \ref{eq:shev5a}		& 2620.5& 1.0			& 183.0		& 			& 			& & \nodata \\
	& 	& IA	& \ref{eq:shev5b} 	& 7891.0& 0.613		& 3.6			& 5.0		& 			& & \nodata \\
17& 19& D		& \ref{eq:shev5a}		& 2818.1& 1.0			& 133.0		& 			& 			& & \nodata \\
	& 	& IA	& \ref{eq:shev5b}		& 7989.0& 0.592		& 3.6			& 5.0		& 			& & \nodata \\
18& 20& D		& \ref{eq:shev5a}		& 3035.9& 1.0			& 76.0		& 			& 			& & \nodata \\
	& 	& IA	& \ref{eq:shev5b}		& 8088.0& 0.565		& 3.6			& 5.0		& 			& & \nodata \\
19& 21& D		& \ref{eq:shev5a}		& 3262.6& 1.0			& 43.0		& 			& 			& & \nodata \\
	& 	& IA	& \ref{eq:shev5b}		& 8187.0& 0.572		& 3.6			& 5.0		& 			& & \nodata \\
20& 22& D		& \ref{eq:shev5a}		& 3485.4& 1.0			& 23.0		& 			& 			& & \nodata \\
	& 	& IA	& \ref{eq:shev5b}		& 8286.0& 0.568		& 3.6			& 5.0		& 			& & \nodata \\
21& 23& D		& \ref{eq:shev5a}		& 3748.8& 1.0			& 10.0		& 			& 			& & \nodata \\
	& 	& IA	& \ref{eq:shev5b}		& 8384.0& 0.729		& 3.6			& 5.0		& 			& & \nodata \\
22& 24& D		& \ref{eq:shev5a}		& 3996.2& 1.0			& 1.8			& 			& 			& & \nodata \\
	& 	& IA	& \ref{eq:shev5b}		& 8482.0& 0.877		& 3.6			& 4.5		& 			& & \nodata \\
23& 25& D		& \ref{eq:shev5a}		& 10847	& 1.0			& 12.0		& 			& 			& & \nodata \\
24& 26& IA	& \ref{eq:shev5a}		& 18106	& 1.0			& 5.0			& 			& 			& & \nodata
\enddata
\tablecomments{The processes D and IA denote direct ionization and ionization autoionization, respectively. EIMI cross sections are derived from the semiempirical formulae and compared to experiment whenever possible. The available experimental data are referenced in the table. The cross sections are discussed in detail in the text.}
\end{deluxetable}

\bibliography{MEII}% Produces the bibliography via BibTeX.

\end{document}